\documentstyle[aps,twocolumn]{revtex}

\begin{document}
\author{Norman Dombey$^{a}$}
\address{Department of Physics and Astronomy\\
University of Sussex\\
Brighton BN1 9QH, UK}
\title{The hydrino and other unlikely states }
\date{August 8, 2006}
\address{$^{a}$normand@sussex.ac.uk,\\
\flushleft We discuss the tightly bound (hydrino) solution of
theKlein-Gordon equation for the Coulomb potential in 3 
dimensions. We show that a similar tightly bound state 
occurs for the Dirac equation for the
Coulomb potential in 2 dimensions. These states are 
unphysical since they disappear if the nuclear charge 
distribution is taken to have an arbitrarily
small but non-zero radius. }
\maketitle

\section{The Hydrino State}

Naudts \cite{Naudts} recently drew attention to the so-called hydrino state
of hydrogen, a strongly bound state which has a normalisable solution of the
Klein-Gordon equation for a Coulomb potential in three space dimensions, but
no analogue in the Schr\"{o}dinger equation. It has been mentioned by a
minority of quantum mechanics textbooks in the past \cite{text}, which
discard it for various {\it ad hoc} reasons: for instance, because it lacks
a nonrelativistic analogue, or because it yields divergent expressions for
the expectation value of the potential energy. Claims that a tightly-bound
state of hydrogen exists have, however, been widely publicised \cite{news}.
Another relativistic equation, the Bethe-Salpeter equation, has been shown
to have anomalous solutions which on closer inspection are unphysical \cite
{cut}.

In this letter, we show that similar anomalous tightly-bound states for a
Coulomb potential also exist in the Dirac equation in two space dimensions.
We claim that all these anomalous states are unphysical by developing an
argument used by Schiff \cite{schiff}, which was first used by Loudon \cite
{loudon} for the one dimensional Schr\"{o}dinger equation with the potential 
$-\alpha /\left| x\right| .$

We shall begin with a brief discussion of the hydrino state for the Coulomb
potential $V(r)=-\alpha /r$ which is taken to be the fourth component of a
four-vector. If the solution $\psi (r,\theta ,\varphi )$ for the
Klein-Gordon equation of a particle of mass $m$ is separated out into its
radial and angular components in the usual way and we consider only states
of angular momentum $l=0,$ then if $f(r)=r\psi $ we have (in natural units
where $\hbar =1=c$) 
\begin{equation}
\frac{d^{2}f}{dr^{2}}+\left[ \left( E+\frac{\alpha }{r}\right)
^{2}-m^{2}\right] f(r)=0.  \label{KG}
\end{equation}
Consider bound states so that energy $E<m$. Write $\kappa =\sqrt{m^{2}-E^{2}}%
,\ z=2\kappa r$ and then

\begin{equation}
f^{\prime \prime }(z)+\left[ -\frac{1}{4}+\frac{\lambda }{z}+\frac{(1/4-\mu
^{2})}{z^{2}}\right] f(z)=0,  \label{W}
\end{equation}
where $\lambda =\alpha E/\kappa $ and $\mu =+\sqrt{1/4-\alpha ^{2}}$. Eq. (%
\ref{W}) is called Whittaker's equation \cite{AS} and the solutions can be
expressed in terms of confluent hypergeometric functions. To see this take 
\begin{equation}
f(z)=z^{\mu +1/2}e^{-z/2}w(z)  \label{efn}
\end{equation}
to obtain the defining confluent hypergeometric or Kummer's equation \cite
{AS}

\begin{equation}
zw^{\prime \prime }+(b-z)w^{\prime }-aw=0;\,a=\mu +1/2-\lambda ,\,b=2\mu +1
\label{confl}
\end{equation}
The roots $t$ of the indicial equation, giving the lowest power of $z$ when $%
w$ is expressed as a power series, are $t=0$ and $t=1-b=-2\mu $. The
solutions of Eq.(\ref{confl}) which allow the wave function $\psi $ to be
square integrable to infinity are polynomials $w(z)=M(a,b,z)$ with $a=-N$
where $N$ is a non-negative integer.

The conventional choice is $t=0$; then $\lambda _{N}=\mu +1/2+N$, whence

\begin{equation}
E_{N}/m=\left[ 1+\frac{\alpha ^{2}}{\left( N+1/2+\sqrt{1/4-\alpha ^{2}}%
\right) ^{2}}\right] ^{-1/2}  \label{nn}
\end{equation}
In particular the ground-state energy $E_{0}$ is given by

\begin{equation}
E_{0}/m=\left( 1/\sqrt{2}\right) \sqrt{1+\sqrt{1-4\alpha ^{2}}}
\label{ground}
\end{equation}
It is easy to show that to lowest order in small $\alpha $ Eq.(\ref{nn})
gives the non-relativistic spectrum for the hydrogen atom obtained from the
Schr\"{o}dinger equation.

\medskip There are also anomalous solutions of Eq. (\ref{W}) from the other
root $t=-2\mu $ (or equivalently set $\mu =-\sqrt{1/4-\alpha ^{2}}$). They
give the spectrum

\[
E_{NA}/m=\left[ 1+\frac{\alpha ^{2}}{\left( N+1/2-\sqrt{1/4-\alpha ^{2}}%
\right) ^{2}}\right] ^{-1/2} 
\]
and in particular

\begin{equation}
E_{0A}/m=\left( 1/\sqrt{2}\right) \sqrt{1-\sqrt{1-4\alpha ^{2}}}
\label{hydr}
\end{equation}
This is the solution which is called the hydrino. It is normalisable but has
no counterpart in the Schrodinger equation. Indeed for small $\alpha ,$ $%
E_{0A}=m\alpha $ and the electron is tightly bound.

It is tempting to say with Migdal \cite{mig} that the one particle
Klein-Gordon equation cannot be taken seriously and field theory is
required. But we now demonstrate similar tightly bound states for small $%
\alpha $ in the one-particle Dirac equation, where tight binding should not
cause a problem\cite{lev}.

\section{The Coulomb Potential in the Two-Dimensional Dirac Equation}

This problem has been discussed previously \cite{2D} and anomalous solutions
were noticed by Sigurdsson \cite{sig}. In two space-dimensions we can take
the gamma matrices $\gamma _{x},\gamma _{y}$ and $\gamma _{0}$ to be the
Pauli matrices $i\sigma _{x},i\sigma _{y}$ and $\sigma _{z}$ respectively.
Then the Dirac equation for a particle of definite energy $E$ in the
presence of the potential $V(r)=-\alpha /r$ is

\begin{equation}
\left[ \sigma _{x}\frac{\partial }{\partial x}+\sigma _{y}\frac{\partial }{%
\partial y}-(E-V)\sigma _{z}+m\right] \psi =0  \label{dir}
\end{equation}
Convert to polar coordinates and take

\begin{equation}
\psi (r,\theta )=\left( 
\begin{array}{c}
f(r)e^{i(j-1/2)\theta } \\ 
g(r)^{i(j+1/2)\theta }
\end{array}
\right)
\end{equation}
which is an eigenstate of the z-component of total angular momentum $J_{z}=$ 
$L_{z}+\frac{1}{2}\sigma _{z}$ with eigenvalue $j$. Eq. (\ref{dir}) now
becomes

\begin{mathletters}
\begin{eqnarray}
\left( m-E-\alpha /r\right) f+dg/dr+\left( j+1/2\right) g/r &=&0 \\
df/dr-\left( j-1/2\right) f/r+(E+\alpha /r+m)g &=&0
\end{eqnarray}
We follow the approach of Tutik \cite{tut}. Write

\begin{center}
$f=F\sqrt{(1+E/m)},\quad g=G\sqrt{(1-E/m)}$
\end{center}

and $\Phi _{1,2}=F\pm G$, and scale to $z$ to obtain

\end{mathletters}
\begin{mathletters}
\begin{eqnarray}
\frac{d\Phi _{1}}{dz}+\frac{\Phi _{1}}{2z}-j\frac{\Phi _{2}}{z}+\frac{1}{2}%
\Phi _{1} &=&\frac{\alpha }{\kappa z}\left( E\Phi _{1}+m\Phi _{2}\right) \\
\frac{d\Phi _{2}}{dz}+\frac{\Phi _{2}}{2z}-j\frac{\Phi _{1}}{z}-\frac{1}{2}%
\Phi _{2} &=&-\frac{\alpha }{\kappa z}\left( m\Phi _{1}+E\Phi _{2}\right)
\end{eqnarray}
Take $\Phi _{1,2}=\phi _{1,2}\exp (-z/2)$; then

\end{mathletters}
\begin{equation}
z\frac{d\phi _{1}}{dz}+\left( \frac{1}{2}-\frac{\alpha E}{\kappa }\right)
\phi _{1}=\left( j+\frac{\alpha m}{\kappa }\right) \phi _{2}
\end{equation}
so $\phi _{2}$ can be written in terms of $\phi _{1\text{. }}$The second
order differential equation for $\phi _{1}$ is

\begin{eqnarray}
&&z\frac{d}{dz}\left[ z\frac{d\phi _{1}}{dz}+\left( \frac{1}{2}-\frac{\alpha
E}{\kappa }\right) \phi _{1}\right] + \\
&&\left( \frac{\alpha E}{\kappa }+\frac{1}{2}-z\right) \left[ z\frac{d\phi
_{1}}{dz}+\left( \frac{1}{2}-\frac{\alpha E}{\kappa }\right) \phi _{1}\right]
\nonumber \\
&=&\left( j^{2}-\frac{\alpha ^{2}m^{2}}{\kappa ^{2}}\right) \phi _{1} 
\nonumber
\end{eqnarray}
or 
\begin{eqnarray}
&&z\phi _{1}^{\prime \prime }(z)+(2-z)\phi _{1}^{\prime }-\left( 1/2-\alpha
E/\kappa \right) \phi _{1}  \nonumber  \label{check} \\
&=&(j^{2}-1/4-\alpha ^{2})\phi _{1}/z
\end{eqnarray}

Write $\phi _{1}=z^{t}w$ so we obtain

\begin{eqnarray}
&&zw^{\prime \prime }(z)+(2+2t-z)w^{\prime }+(t^{2}+t)w/z-tw \\
&=&\left( \frac{j^{2}-1/4-\alpha ^{2}}{z}+\frac{1}{2}-\frac{\alpha E}{\kappa 
}\right) w
\end{eqnarray}
Hence take

\[
t^{2}+t=j^{2}-1/4-\alpha ^{2} 
\]
or

\begin{equation}
t=-1/2\pm \sqrt{j^{2}-\alpha ^{2}}  \label{choose}
\end{equation}
to obtain

\begin{equation}
zw^{\prime \prime }(z)+(2+2t-z)w^{\prime }-\left( t+\frac{1}{2}-\frac{\alpha
E}{\kappa }\right) w=0
\end{equation}
which again is Kummer's equation. The regular solution $M(a,b,z)$ is square
integrable to infinity only if it reduces to a polynomial, which it does if

\begin{equation}
a=t+1/2-\alpha E/\kappa =-N  \label{check3}
\end{equation}
where $N$ is a non-negative integer.

Conventionally one chooses the positive sign in Eq. (\ref{choose}) which
yields

\begin{equation}
\alpha E_{N}/\kappa =N+\sqrt{j^{2}-\alpha ^{2}}
\end{equation}
or for $j=\frac{1}{2}$

\begin{equation}
\frac{E_{N}}{m}=\left[ 1+\frac{\alpha ^{2}}{(N+\sqrt{1/4-\alpha ^{2}})^{2}}%
\right] ^{-1/2}  \label{rel}
\end{equation}
and the ground state is given by

\begin{equation}
E_{0}/m=\sqrt{1-4\alpha ^{2}}
\end{equation}
As $r\rightarrow 0,$ $f,g\sim r^{t}=r^{-\frac{1}{2}+\sqrt{\frac{1}{4}-\alpha
^{2}}\text{ }}$so that the wave functions remain square-integrable down to
the origin even though they diverge. They are the relativistic
generalisation of the solutions of the Schr\"{o}dinger equation in two
dimensions which has the energy spectrum given by Eq. (\ref{rel}) to lowest
order in $\alpha ^{2}.$

The anomalous solutions arise from the other root of Eq. (\ref{choose}): $%
t=-1/2-\sqrt{1/4-\alpha ^{2}}.$ They too are square integrable down to the
origin. The anomalous energy eigenvalues are given by $a=t+1/2-\alpha
E_{A}/\kappa =-N$ or

\begin{equation}
\alpha E_{NA}/\kappa =N-\sqrt{1/4-\alpha ^{2}}
\end{equation}
Because $\kappa >0$, the lowest anomalous state with $N=0$ satisfies

\[
\alpha E_{0A}/\kappa =-\sqrt{1/4-\alpha ^{2}} 
\]
whence

\begin{equation}
E_{0A}/m=-\sqrt{1-4\alpha ^{2}}
\end{equation}
This solution has no non-relativistic analogue and like the hydrino
corresponds to a state which is tightly bound for small $\alpha $.

\section{Discussion}

We believe that the three dimensional Klein-Gordon hydrino state and its two
dimensional Dirac analogue are unphysical in spite of having normalisable
wave functions. Their obvious failings are that

\begin{enumerate}
\item  they lack non-relativistic counterparts even for arbitrarily small
coupling;

\item  the states persist even when the coupling is turned off;

\item  the strength of the binding increases as the coupling strength $%
\alpha $ decreases. The maximum binding occurs for $\alpha =0$ when the
potential has disappeared completely\footnote{%
We could call these anomalous states homeopathic states because the smaller
the coupling, the larger the effect.}.
\end{enumerate}

These reasons may be sound but they are not decisive either mathematically
or physically. We now demonstrate that, if the point charge of the nucleus
is replaced by a charge extending over an arbitrarily small but finite
radius $R$, then the anomalous \ functions become unacceptable because for
small enough $R$ they cease to satisfy the appropriate wave equation. We
will prove this for the Klein-Gordon equation; the argument for the Dirac
case runs similarly.

For simplicity, we look at a conducting spherical shell of radius $r=R$, so
that $V(r<R)=-\alpha /R$ while $V(r>R)=-\alpha /r$. We need consider only $%
\alpha ^{2}\ll 1$ since we are investigating the bound-state energy in the
limit $\alpha \rightarrow 0$. Then the inner solution of Eq. (\ref{KG}) is

\begin{equation}
f(r<R)=C\sin (qr),\quad q=\sqrt{(E+\alpha /R)^{2}-m^{2}}
\end{equation}
For $R$ small enough to make $\alpha /R>>m$, one has $\quad q\rightarrow
\alpha /R$, so that $qr=\alpha r/R\leq \alpha \ll 1\Rightarrow $ $%
f(r)=Cqr(1+O((qr)^{2})$ .\ In terms of the scaled variable $z$, and defining 
$Z\equiv 2\kappa R$, this entails

\begin{equation}
\left. \frac{zdf(z)/dz}{f(z)}\right| _{z=Z}=1+O(Z)  \label{match}
\end{equation}

This solution has to be joined to the outer solution of Eq. (\ref{W}) for $%
z>Z=2\kappa R$. For $w(z)$ defined by (\ref{wf}) the general solution of Eq.
(\ref{confl}) reads\cite{AS} 
\begin{equation}
w=AM(a,b,z)+Bz^{1-b}M(a-b+1,2-b,z)
\end{equation}
where as $z\rightarrow \infty $

\begin{equation}
M(c,d,z\rightarrow \infty )\rightarrow \frac{\Gamma (d)}{\Gamma (c)}\exp
(z)z^{c-d}  \label{far}
\end{equation}
Hence 
\begin{equation}
w(z\rightarrow \infty )\rightarrow \left\{ A\frac{\Gamma (b)}{\Gamma (a)}+B%
\frac{\Gamma (2-b)}{\Gamma (a-b+1)}\right\} \exp (z)z^{a-b}.  \label{FAR}
\end{equation}

For a bound state the expression in the brackets must vanish so that 
\begin{eqnarray}
w &=&M(a,b,z)-  \nonumber \\
&&\frac{\Gamma (b)\Gamma (a-b+1)}{\Gamma (a)\Gamma (2-b)}%
z^{1-b}M(a-b+1,2-b,z)
\end{eqnarray}
Thus for small $z$

\begin{eqnarray}
w(z &\rightarrow &0)\rightarrow 1-  \nonumber \\
&&\frac{\Gamma (b)\Gamma (a-b+1)}{\Gamma (a)\Gamma (2-b)}\left\{ z^{1-b}+%
\frac{(a-b+1)}{(2-b)}z^{2-b}+...\right\}
\end{eqnarray}
hence 
\begin{equation}
f(z\rightarrow 0)\rightarrow (1-z/2+...)z^{\mu +1/2}w
\end{equation}
Recall $a=\mu +1/2-\alpha E/\kappa $, $b=2\mu +1$, $\mu =\sqrt{1/4-\alpha
^{2}}$, so for small $\alpha $%
\[
1-b\simeq -1+2\alpha ^{2},\qquad 2-b\simeq 2\alpha ^{2}. 
\]
Since we are considering the limit as $\alpha \rightarrow 0,$ we can
disregard $\alpha ^{2}$ in the various exponents of $z$ and so as $%
z\rightarrow 0$%
\begin{equation}
f=z-\frac{\Gamma (b)\Gamma (a-b+1)}{\Gamma (a)\Gamma (2-b)}(1+z+...)
\end{equation}
Hence requiring $f(z)$ to be proportional to $z$ for small $z$ in order to
satisfy Eq. (\ref{match})gives

\begin{equation}
1/\Gamma (a)=0\qquad \Rightarrow \qquad a=-N,\text{\qquad }N=0,1,2,...
\end{equation}
This is precisely the conventional quantisation condition which gives the
energy eigenvalue spectrum of Eq. (\ref{nn}) thus eliminating the anomalous
solutions.

For the two dimensional Dirac equation for a Coulomb potential a similar
argument, with Bessel functions replacing trigonometric functions for the
inner solution, will give $f(r=R)\rightarrow $ const; $g(r=R)\rightarrow 0$
for $R$ small enough and hence $\phi _{1}=z^{t}w$ $\rightarrow $ const. This
is only possible if the positive sign is taken in Eq. (\ref{choose}), ruling
out the anomalous solutions here too.

\section{Conclusions}

We have shown that there are normalisable solutions of the Dirac equation
for the Coulomb potential in two dimensions which are tightly bound, just
like the hydrino solution of the Klein-Gordon equation in three dimensions.
We have displayed the unphysical nature of such solutions by smearing the
nuclear charge distribution so that it has an arbitrarily small non-zero
radius; in that case there are no anomalous solutions. We suggest that
outside of science fiction this is sufficient reason to disregard them. We
leave to others to find a general criterion to exclude the anomalous
solutions for the Coulomb potential with a point charge in two and three
space dimensions: in the Appendix we indicate a possible route to this end.

{\bf Acknowledgements} I would like to thank Gabriel Barton for suggesting
that the Coulomb potential should be cut off at small distances and showing
me how to manipulate the confluent hypergeometric functions. I would also
like to thank the referee for suggesting the insertion of the Appendix.

\section{APPENDIX}

The requirement that wave functions have to be square integrable is not the
only restriction on physical states in quantum mechanics. As de Castro has
pointed out in his study of Klein Gordon particles \cite{deC} for a similar
problem, it is also necessary to ensure that the Hamiltonian is Hermitian.
We shall not attempt to do this here, but a necessary consequence of
Hermiticity will be that the eigenfunctions of the Hamiltonian corresponding
to two distinct energy eigenvalues are orthogonal.

The eigenfunctions $f_{0}(z)$ for the normal ground state energy given by
Eq. (\ref{ground}) and $f_{0A}(z)$ for the hydrino energy given by Eq. (\ref
{hydr}) are similar. They are both of the form given by Eq. (\ref{efn}):

\begin{equation}
f(z)=z^{\mu +1/2}e^{-z/2}w(z)
\end{equation}
where in both cases $w(z)=M(a,b,z)$ and $a=0$; i.e. $w(z)$ is constant.
Neither $f_{0}(z)$ nor $f_{0A}(z)$ vanishes except at $z=0$. Hence the
hydrino wave function $\psi _{0A}=f_{0A}(z)/r$ cannot be orthogonal to $\psi
_{0}=f_{0}(z)/r.$ Thus the hydrino state cannot be physical provided that it
is possible to ensure that the Hamiltonian for the hydrogen atom in the
Klein-Gordon equation is Hermitian.

\end{document}